\errorstopmode
\input amssym.def
\input amssym.tex


\magnification=\magstephalf
\hsize=14.0 true cm
\vsize=19 true cm
\hoffset=1.0 true cm
\voffset=2.0 true cm

\abovedisplayskip=12pt plus 3pt minus 3pt
\belowdisplayskip=12pt plus 3pt minus 3pt
\parindent=1.0em


\font\sixrm=cmr6
\font\eightrm=cmr8
\font\ninerm=cmr9

\font\sixi=cmmi6
\font\eighti=cmmi8
\font\ninei=cmmi9

\font\sixsy=cmsy6
\font\eightsy=cmsy8
\font\ninesy=cmsy9

\font\sixbf=cmbx6
\font\eightbf=cmbx8
\font\ninebf=cmbx9

\font\eightit=cmti8
\font\nineit=cmti9

\font\eightsl=cmsl8
\font\ninesl=cmsl9

\font\sixss=cmss8 at 8 true pt
\font\sevenss=cmss9 at 9 true pt
\font\eightss=cmss8
\font\niness=cmss9
\font\tenss=cmss10

\font\sixmib=cmmib6
\font\sevenmib=cmmib7
\font\eightmib=cmmib8
\font\ninemib=cmmib9
\font\tenmib=cmmib10

 at 12 true pt
 at 12 true pt
\font\bigrm=cmr10 at 12 true pt
 at 12 true pt
 at 12 true pt

 at 16 true pt
 at 16 true pt
\font\Bigrm=cmr12 at 16 true pt
 at 16 true pt
 at 16 true pt

\catcode`@=11
\newfam\ssfam
\newfam\mibfam

\def\tenpoint{\def\rm{\fam0\tenrm}%
    \textfont0=\tenrm \scriptfont0=\sevenrm \scriptscriptfont0=\fiverm
    \textfont1=\teni  \scriptfont1=\seveni  \scriptscriptfont1=\fivei
    \textfont2=\tensy \scriptfont2=\sevensy \scriptscriptfont2=\fivesy
    \textfont3=\tenex \scriptfont3=\tenex   \scriptscriptfont3=\tenex
    \textfont\itfam=\tenit                  \def\it{\fam\itfam\tenit}%
    \textfont\slfam=\tensl                  \def\sl{\fam\slfam\tensl}%
    \textfont\bffam=\tenbf \scriptfont\bffam=\sevenbf
                           \scriptscriptfont\bffam=\fivebf
                           \def\bf{\fam\bffam\tenbf}%
    \textfont\ssfam=\tenss \scriptfont\ssfam=\sevenss
                           \scriptscriptfont\ssfam=\sevenss
                           \def\ss{\fam\ssfam\tenss}%
    \textfont\mibfam=\tenmib \scriptfont\mibfam=\sevenmib
                             \scriptscriptfont\mibfam=\sevenmib
                             \def\mib{\fam\mibfam\tenmib}%
    \normalbaselineskip=13pt
    \setbox\strutbox=\hbox{\vrule height8.5pt depth3.5pt width0pt}%
    \let\big=\tenbig
    \normalbaselines\rm}

\def\ninepoint{\def\rm{\fam0\ninerm}%
    \textfont0=\ninerm      \scriptfont0=\sixrm
                            \scriptscriptfont0=\fiverm
    \textfont1=\ninei       \scriptfont1=\sixi
                            \scriptscriptfont1=\fivei
    \textfont2=\ninesy      \scriptfont2=\sixsy
                            \scriptscriptfont2=\fivesy
    \textfont3=\tenex       \scriptfont3=\tenex
                            \scriptscriptfont3=\tenex
    \textfont\itfam=\nineit \def\it{\fam\itfam\nineit}%
    \textfont\slfam=\ninesl \def\sl{\fam\slfam\ninesl}%
    \textfont\bffam=\ninebf \scriptfont\bffam=\sixbf
                            \scriptscriptfont\bffam=\fivebf
                            \def\bf{\fam\bffam\ninebf}%
    \textfont\ssfam=\niness \scriptfont\ssfam=\sixss
                            \scriptscriptfont\ssfam=\sixss
                            \def\ss{\fam\ssfam\niness}%
    \textfont\mibfam=\ninemib \scriptfont\mibfam=\sixmib
                            \scriptscriptfont\mibfam=\sixmib
                            \def\mib{\fam\mibfam\ninemib}%
    \normalbaselineskip=12pt
    \setbox\strutbox=\hbox{\vrule height8.0pt depth3.0pt width0pt}%
    \let\big=\ninebig
    \normalbaselines\rm}

\def\eightpoint{\def\rm{\fam0\eightrm}%
    \textfont0=\eightrm      \scriptfont0=\sixrm
                             \scriptscriptfont0=\fiverm
    \textfont1=\eighti       \scriptfont1=\sixi
                             \scriptscriptfont1=\fivei
    \textfont2=\eightsy      \scriptfont2=\sixsy
                             \scriptscriptfont2=\fivesy
    \textfont3=\tenex        \scriptfont3=\tenex
                             \scriptscriptfont3=\tenex
    \textfont\itfam=\eightit \def\it{\fam\itfam\eightit}%
    \textfont\slfam=\eightsl \def\sl{\fam\slfam\eightsl}%
    \textfont\bffam=\eightbf \scriptfont\bffam=\sixbf
                             \scriptscriptfont\bffam=\fivebf
                             \def\bf{\fam\bffam\eightbf}%
    \textfont\ssfam=\eightss \scriptfont\ssfam=\sixss
                             \scriptscriptfont\ssfam=\sixss
                             \def\ss{\fam\ssfam\eightss}%
    \textfont\mibfam=\eightmib \scriptfont\mibfam=\sixmib
                             \scriptscriptfont\mibfam=\sixmib
                             \def\mib{\fam\mibfam\eightmib}%
    \normalbaselineskip=10pt
    \setbox\strutbox=\hbox{\vrule height7.0pt depth2.0pt width0pt}%
    \let\big=\eightbig
    \normalbaselines\rm}

\def\tenbig#1{{\hbox{$\left#1\vbox to8.5pt{}\right.\n@space$}}}
\def\ninebig#1{{\hbox{$\textfont0=\tenrm\textfont2=\tensy
                       \left#1\vbox to7.25pt{}\right.\n@space$}}}
\def\eightbig#1{{\hbox{$\textfont0=\ninerm\textfont2=\ninesy
                       \left#1\vbox to6.5pt{}\right.\n@space$}}}

\font\sectionfont=cmbx10
\font\subsectionfont=cmti10

\def\figurecaptionfont{\ninepoint}
\def\tablecaptionfont{\ninepoint}
\def\footnotefont{\eightpoint}


\newcount\equationno
\newcount\bibitemno
\newcount\figureno
\newcount\tableno

\equationno=0
\bibitemno=0
\figureno=0
\tableno=0


\footline={\ifnum\pageno=0{\hfil}\else
{\hss\rm\the\pageno\hss}\fi}


\def\section #1. #2 \par
{\vskip0pt plus .10\vsize\penalty-100 \vskip0pt plus-.10\vsize
\vskip 1.6 true cm plus 0.2 true cm minus 0.2 true cm
\global\def\equationlabel{#1}
\global\equationno=0
\leftline{\sectionfont #1. #2}\par
\immediate\write\terminal{Section #1. #2}
\vskip 0.7 true cm plus 0.1 true cm minus 0.1 true cm
\noindent}


\def\subsection #1 \par
{\vskip0pt plus 0.8 true cm\penalty-50 \vskip0pt plus-0.8 true cm
\vskip2.5ex plus 0.1ex minus 0.1ex
\leftline{\subsectionfont #1}\par
\immediate\write\terminal{Subsection #1}
\vskip1.0ex plus 0.1ex minus 0.1ex
\noindent}


\def\appendix #1. #2 \par
{\vskip0pt plus .20\vsize\penalty-100 \vskip0pt plus-.20\vsize
\vskip 1.6 true cm plus 0.2 true cm minus 0.2 true cm
\global\def\equationlabel{\hbox{\rm#1}}
\global\equationno=0
\leftline{\sectionfont Appendix #1. #2}\par
\immediate\write\terminal{Appendix #1. #2}
\vskip 0.7 true cm plus 0.1 true cm minus 0.1 true cm
\noindent}



\def\equation#1{$$\displaylines{\qquad #1}$$}
\def\enum{\global\advance\equationno by 1
\hfill\llap{{\rm(\equationlabel.\the\equationno)}}}

\def\next#1{\cr\noalign{\vskip#1}\qquad}


\def\ifundefined#1{\expandafter\ifx\csname#1\endcsname\relax}

\def\ref#1{\ifundefined{#1}?\immediate\write\terminal{unknown reference
on page \the\pageno}\else\csname#1\endcsname\fi}

\newwrite\terminal
\newwrite\bibitemlist

\def\bibitem#1#2\par{\global\advance\bibitemno by 1
\immediate\write\bibitemlist{\string\def
\expandafter\string\csname#1\endcsname
{\the\bibitemno}}
\item{[\the\bibitemno]}#2\par}

\def\beginbibliography{
\vskip0pt plus .15\vsize\penalty-100 \vskip0pt plus-.15\vsize
\vskip 1.2 true cm plus 0.2 true cm minus 0.2 true cm
\leftline{\sectionfont References}\par
\immediate\write\terminal{References}
\immediate\openout\bibitemlist=biblist
\frenchspacing\parindent=1.8em
\vskip 0.5 true cm plus 0.1 true cm minus 0.1 true cm}

\def\endbibliography{
\immediate\closeout\bibitemlist
\nonfrenchspacing\parindent=1.0em}


\def\figurecaption#1{\global\advance\figureno by 1
\narrower\figurecaptionfont
Fig.~\the\figureno. #1}

\def\tablecaption#1{\global\advance\tableno by 1
\vbox to 0.25 true cm { }
\centerline{\tablecaptionfont%
Table~\the\tableno. #1}
\vskip-0.4 true cm}

\def\thicktablerule{\hrule height1pt}
\def\thintablerule{\hrule height0.4pt}

\tenpoint

\immediate\openin\bibitemlist=biblist
\ifeof\bibitemlist\immediate\closein\bibitemlist
\else\immediate\closein\bibitemlist
\input biblist \fi


\def\thismonth{\ifcase\month\or
January\or February\or March\or April\or May\or June\or
July\or August\or September\or October\or November\or December\fi}

\input epsf
\epsfclipon



\def\rme{{\rm e}}
\def\rmO{{\rm O}}



\def\proof{\noindent{\sl Proof:}\kern0.6em}

\def\frac#1#2{\hbox{$#1\over#2$}}
\def\dual{\mathstrut^*\kern-0.1em}

\def\lvec#1{\setbox0=\hbox{$#1$}
    \setbox1=\hbox{$\scriptstyle\leftarrow$}
    #1\kern-\wd0\smash{
    \raise\ht0\hbox{$\raise1pt\hbox{$\scriptstyle\leftarrow$}$}}
    \kern-\wd1\kern\wd0}
\def\rvec#1{\setbox0=\hbox{$#1$}
    \setbox1=\hbox{$\scriptstyle\rightarrow$}
    #1\kern-\wd0\smash{
    \raise\ht0\hbox{$\raise1pt\hbox{$\scriptstyle\rightarrow$}$}}
    \kern-\wd1\kern\wd0}
\def\slash#1{\setbox0=\hbox{$#1$}\setbox1=\hbox{$\kern1pt/$}
    #1\kern-\wd0\kern1pt/\kern-\wd1\kern\wd0}


\def\nabstar#1{{\nabla\kern0.5pt\smash{\raise 4.5pt\hbox{$\ast$}}
               \kern-5.5pt_{#1}}}

\def\drvstar#1{{\partial\kern0.5pt\smash{\raise 4.5pt\hbox{$\ast$}}
               \kern-6.0pt_{#1}}}

\def\ldrvstar#1{{\lvec{\,\partial}\kern-0.5pt\smash{\raise 4.5pt\hbox{$\ast$}}
               \kern-5.0pt_{#1}}}


\def\MeV{{\rm MeV}}

\def\fm{{\rm fm}}



\def\dmode{\phi}


\def\diracstar#1#2{
    \setbox0=\hbox{$\gamma$}\setbox1=\hbox{$\gamma_{#1}$}
    \gamma_{#1}\kern-\wd1\kern\wd0
    \smash{\raise4.5pt\hbox{$\scriptstyle#2$}}}


\def\Ad{{\rm Ad}\kern0.1em}


\def\ms{m_s}

\def\msea{m_{\rm sea}}
\def\mval{m_{\rm val}}
\def\ZA{Z_{\rm A}}
\def\ZP{Z_{\rm P}}
\def\ksea{\kappa_{\rm sea}}
\def\kval{\kappa_{\rm val}}


\def\Dhat{\hat{D}}
\def\csw{c_{\rm sw}}

\def\Zop{A}
\def\ZopB{B}
\def\subsp{{\cal S}}
\def\Nkv{N_{\rm kv}}
\def\Nm{N_{s}}

%
\rightline{CERN-PH-TH/2007-096}

\vskip0.8cm 
\centerline{\Bigrm
Local coherence and deflation of the low
}
\vskip0.2cm
\centerline{\Bigrm
quark modes in lattice QCD
}
\vskip 0.6 true cm
\centerline{\bigrm Martin L\"uscher}
\vskip1ex
\centerline{\it CERN, Physics Department, TH Division}
\centerline{\it CH-1211 Geneva 23, Switzerland}
\vskip 0.8 true cm
\thintablerule
\vskip 2.0ex
\ninepoint
\leftline{\bf Abstract}
\vskip 1.0ex\noindent
The spontaneous breaking of chiral symmetry in QCD 
is known to be linked to a non-zero density
of eigenvalues of the massless Dirac operator near the origin.
Numerical studies of two-flavour QCD now suggest
that the low quark modes are locally coherent
to a certain extent. As a consequence,
the modes can be simultaneously 
deflated, using local projectors, with a total
computational effort proportional to the lattice volume
(rather than its square). Deflation has potentially many uses
in lattice QCD. The technique is here worked out for the case
of quark propagator calculations, where large speed-up factors
and a flat scaling behaviour with respect to the quark mass
are achieved.
\vskip 2.0ex
\thintablerule

\tenpoint

\vskip-0.3cm

\section 1. Introduction

The physical masses of the up and down quarks are 
much smaller than the typical low-energy hadronic scales
such as the pion decay constant and the string tension.
In numerical lattice QCD, the smallness of the quark
masses still is a source of difficulty, for various
reasons, but mainly because the available simulation techniques
become inefficient close to the chiral limit.

It is not excluded, however,
that many of the present limitations in lattice QCD can
be overcome by ``deflating QCD'', i.e.~by treating
the eigenmodes of the Dirac operator with small eigenvalues 
separately from the bulk of the quark modes.
Deflation techniques are used in many areas of applied science
and they are also an active research topic in numerical mathematics
(see refs.~[\ref{FrankVuik},\ref{NabbenVuik}], for example, and
references quoted there). In lattice QCD low-mode deflation was so far 
mainly used in connection with statistical error reduction methods 
[\ref{NeffEtAl}--\ref{AllToAllTrinity}] that now go under the headings of
{\it low-mode averaging} and {\it all-to-all propagators}.
Other applications of deflation methods in QCD include
quark propagator computations in special situations, where 
only a small number of modes need to be deflated
[\ref{deForcrand}--\ref{LowModeKpipi}].

In the large-volume regime of QCD, the 
low-mode deflation methods proposed to date
however tend to become useless in practice,
because the number of eigenvalues of the Dirac operator
below any fixed value, say $100$ MeV, grows proportionally
to the four-dimensional volume $V$ of the lattice.
The computational effort required for the calculation of 
the low quark modes and the deflation operations scales like $V^2$ 
in this situation (or even a higher power of $V$) and eventually
offsets the benefits of low-mode deflation.
As Banks and Casher [\ref{BanksCasher}] noted long ago,
the average spectral density of the low quark modes is proportional to the
quark condensate in the chiral limit.
The $V^2$--problem is thus directly linked to 
the spontaneous breaking of chiral symmetry and 
is therefore present independently
of the chosen lattice formulation of the theory.

At present little appears to be known about the 
space-time structure of the low quark modes,
but a simple numerical inspection, reported in sect.~5, 
suggests that they are locally coherent to some extent.
This property allows highly effective 
deflation subspaces to be built from only a few low modes,
using block projectors.
The numerical effort required for the preparation of the subspace and 
the deflation of the Dirac operator is then only of order $V$
(rather than $V^2$).

Before going into the details of the construction in sects.~4 and 5, 
the practical relevance of the $V^2$--problem 
is briefly discussed in sect.~2 and it
is explained, in sect.~3, how to deflate the Dirac operator if
the deflation subspace is not spanned by exact 
eigenmodes of the operator.
The potential
of the proposed deflation method is demonstrated in sect.~6, 
where a preconditioned solver for the lattice Dirac
equation is described,
whose efficiency decreases only slightly with the quark mass
and which outperforms any solver previously used in lattice QCD 
by a large factor.

\section 2. Spectral density and the $V^2$--problem 

\vskip-2.5ex plus 1.0ex

\subsection 2.1 Lattice parameters \& field ensembles

All simulation results reported in this paper were obtained using
the O($a$)-improved Wilson formulation of lattice QCD
[\ref{SW},\ref{OaImp}] with two flavours of mass-degenerate sea quarks.
Only two lattices, of size $48\times24^3$
and $64\times32^3$, were considered, both
at the same inverse gauge coupling $\beta=5.3$, sea-quark hopping 
parameter $\ksea=0.13625$ and value $\csw=1.90952$ [\ref{NPimp}] of the 
Sheikholeslami--Wohlert improvement term. 
At this point in parameter space,
the lattice spacing $a$ in physical units
is estimated to be
$0.0784(10)$ fm [\ref{QCDlite}], while
the sea-quark mass is roughly equal to 
a quarter of the physical strange-quark mass $\ms$.

Representative ensembles of gauge-field configurations
on these two lattices were generated 
by the authors of ref.~[\ref{QCDlite}] and were 
made available for the studies conducted here.
The ensembles consist of $169$ and $50$ configurations, 
widely separated in simulation time so that the
residual autocorrelations can, in most cases, be expected to be negligible.
However, the discussion that follows is intended to be largely
qualitative and the quoted errors and any systematic uncertainties
will therefore be generously ignored.

\subsection 2.2 Computation of the spectral density

In the Wilson theory, the spectrum
of the (massive) lattice Dirac operator $D$ 
is supported in an elliptic region in the complex plane
and is thus not easily compared with the spectrum of the 
Dirac operator in the continuum theory and the 
Banks--Casher formula.
This difficulty can be bypassed by 
considering the hermitian operator $D^{\dagger} D$ instead of $D$,
a choice which has other advantages as well.
The computation of the low-lying
eigenvalues of the operator, for example, 
becomes relatively straightforward. In this paper all 
eigenvalue and eigenmode calculations were performed using 
Chebyshev-accelerated subspace iterations
(see appendix A of ref.~[\ref{Stability}]).

The spectral density of $(D^{\dagger} D)^{1/2}$,
averaged over the ensemble of gauge-field configurations
on the $48\times24^3$ lattice, is shown in fig.~1.
Perhaps the most interesting feature of this distribution is
that it is practically constant 
above the threshold region at the low end of the spectrum.
The threshold of the density in infinite volume is, incidentally, 
expected to be
at $\ZA\msea$ [\ref{Stability}], where
$\ZA$ and $\msea$ denote the axial current renormalization constant
and the bare current-quark mass of the sea quark
($\ZA=0.75(1)$ [\ref{ZAalpha}] and $a\msea=0.00761(7)$ [\ref{QCDlite}]
on the lattices considered here). As can be seen from the figure,
this value appears to give a good indication on where the bulk of the 
spectrum in finite volume begins.

\topinsert
\vbox{
\vskip0.0cm
\epsfxsize=10.0cm\hskip1.0cm\epsfbox{plots/density.eps}
\vskip0.3cm
\figurecaption{%
Unrenormalized density $\rho(\lambda)$
of the eigenvalues $\lambda$ of $(D^{\dagger}D)^{1/2}$
on the $48\times24^3$ lattice, in 
units of `number of eigenvalues per MeV and $\fm^4$'.
The lattice parameters are as specified in subsect.~2.1 
and the dotted vertical line indicates the theoretically expected
position of the threshold of the density
in infinite volume [\ref{Stability}].
}
\vskip0.0cm
}
\endinsert

As discussed in ref.~[\ref{Stability}], the 
spectral density of $(D^{\dagger} D)^{1/2}$
renormalizes multiplicatively,
the renormalization factor $\ZP$ being the same as the one of the 
pseudo-scalar density. For the specified lattice parameters,
the conversion factor from the lattice to the $\overline{\rm MS}$ scheme
of dimensional regularization at renormalization scale $\mu=2$ GeV
was recently determined to be $\ZP^{-1}=1.84(3)$ [\ref{MsALPHA}].
The range of eigenvalues in fig.~1 thus extends up to 
about $121$ MeV after conversion to the $\overline{\rm MS}$ scheme,
i.e.~to a value approximately $25\%$ larger than the 
physical mass of the 
strange quark [\ref{MsJLQCD}--\ref{MsALPHA}].

The spectral density on the $64\times32^3$ lattice was also computed
and turned out to be nearly the same as the one on the $48\times24^3$ lattice.
In particular, the average number of eigenmodes in the 
range covered by fig.~1 increases from $29$ on the 
smaller lattice to about $89$ on the big lattice, which
shows that the $V^2$--problem is not an academic one.
The computation of the $32$ lowest eigenvalues and associated
eigenmodes 
of $D^{\dagger}D$ on the $48\times24^3$ lattice, for example,
to a relative precision of $10^{-3}$,
is in fact already a heavy task
that requires the Dirac operator to be applied some
$2.5\times10^5$ times.

\subsection 2.3 Comparison with the Banks--Casher formula

According to the Banks--Casher relation [\ref{BanksCasher}],
the average number $n(M)$ of eigenvalues 
of the massless Dirac operator of magnitude less than $M$ is,
in the continuum theory, given by
\equation{
  n(M)={2\over\pi}M\Sigma V+\rmO(M^2),
  \enum
}
where $\Sigma$ denotes the $u$-quark condensate in the 
thermodynamic limit. This formula holds in any
renormalization scheme, but $\Sigma$ must refer to a definite 
normalization prescription.
A recently quoted result in two-flavour QCD for the condensate
in the $\overline{\rm MS}$ scheme is $\Sigma=(251\pm13\,\MeV)^3$
[\ref{FukayaEtAl}].
Setting $M=100$ MeV for illustration, and assuming a
$2L\times L^3$ lattice, eq.~(2.1) then yields
the estimates $n(M)=21$, $108$ and $342$ for the average number of 
quark modes below $M$ at $L=2$, $3$ and $4$ fm.

These figures are in a similar range as the
numerically determined mode numbers reported
in the previous subsection. A quantitative comparison must however
take into account the exact physical 
sizes of the simulated lattices and the fact
that the lattice Dirac operator $D$ includes the quark mass term.
Inserting again the value of $\Sigma$ quoted above
and converting the lattice sizes
to physical units (using $a=0.0784$ fm), the Banks--Casher formula
then predicts the average number of eigenvalues in the range
covered by fig.~1 to be $20$ and 
$63$, respectively, on the $48\times24^3$ and the $64\times32^3$ lattice.
These values are smaller 
than the actual numbers ($29$ and $89$) of low modes on these lattices,
but they are in the same ballpark and one should also not forget
that there are systematic uncertainties in these calculations.

\section 3. Inexact deflation

It should be quite clear at this point
that good deflation
methods in QCD should not assume
the low eigenmodes of the Dirac operator to be accurately known.
Eventually the only requirements are that the method
is efficient and that the correctness of the final results 
is guaranteed.
Inexact deflation was already discussed in ref.~[\ref{LowModeKpipi}],
for example,
and will be driven to the extreme in this paper, 
partly following recent developments in the mathematical
literature [\ref{FrankVuik},\ref{NabbenVuik}].

\subsection 3.1 Oblique projector algebra

Deflation methods in QCD usually start from a set 
of quark fields, $\dmode_1(x),\ldots,\dmode_N(x)$,
which will here be assumed to be orthonormal
but are otherwise left unspecified\kern1.5pt\footnote{$\dagger$}{\footnotefont%
The term {\it quark field}\/ is reserved for lattice Dirac fields that 
carry a colour but no flavour index. The eigenmodes of
$D^{\dagger}D$ with small eigenvalues are referred to as the 
{\it low quark modes}\/ or, somewhat abusively, as the 
{\it low modes of the Dirac operator}.}.
The orthogonal projector $P$ to the space $\subsp$ spanned by
these fields (the {\it deflation subspace}) acts on
a given quark field $\psi(x)$ according to
\equation{
   P\psi(x)=\sum_{k=1}^N\dmode_k(x)\left(\dmode_k,\psi\right),
   \enum
}
where the bracket $(\chi,\psi)$ denotes the obvious
scalar product in the linear space of all quark fields.

The restriction of the lattice Dirac operator $D$ 
to the deflation subspace is referred to as the {\it little Dirac operator}.
It is completely specified by the matrix
\equation{
  \Zop_{kl}=\left(\dmode_k,D\dmode_l\right),
  \quad
  k,l=1,\ldots,N,
  \enum
}
that represents its action on the basis fields.
In the following, the little Dirac operator is assumed to be 
invertible, a requirement that will always be satisfied in practice. 
The linear operators
\equation{
   P_L\psi(x)=\psi(x)-\sum_{k,l=1}^ND\dmode_k(x)
   (\Zop^{-1})_{kl}\left(\dmode_l,\psi\right),
   \enum
   \next{2ex}
   P_R\psi(x)=\psi(x)-\sum_{k,l=1}^N\dmode_k(x)
   (\Zop^{-1})_{kl}\left(\dmode_l,D\psi\right),
   \enum
}
can then be defined, where the subscripts stand for ``left'' and 
``right'' because $P_L$ and $P_R$ usually appear on the left and
right of the Dirac operator. These operators are oblique projectors,
i.e.~they are not hermitian but satisfy
\equation{
  P_L^2=P_L,\qquad P_R^2=P_R.
  \enum
}
Other algebraic identities that follow directly from the definitions
(3.1)--(3.4) are 
\equation{
  P_LD=DP_R,
  \enum
  \next{1.5ex}
  PP_L=P_RP=0,
  \enum
  \next{1.5ex}
  P_L(1-P)=(1-P)P_R=1-P.
  \enum
}
In particular, $P_L$ projects to the orthogonal complement
of the deflation subspace.

\subsection 3.2 Deflation of the Dirac equation

The inhomogeneous Dirac equation,
\equation{
  D\psi(x)=\eta(x),
  \enum
}
may now be split into two independent equations by acting with
the projectors $P_L$ and $1-P_L$ from the left. The second equation
can be solved immediately and the solution of the full system
is then given by
\equation{
  \psi(x)=\chi(x)+\sum_{k,l=1}^N\dmode_k(x)(\Zop^{-1})_{kl}\left(\dmode_l,\eta\right),
  \enum
}
where $\chi(x)$ must solve the deflated system
\equation{
  P_LD\chi(x)=P_L\eta(x)
  \enum
}
subject to the constraint $(1-P_R)\chi(x)=0$. 
In view of the commutator property (3.6), this constraint is
consistent with the deflated system and can be freely imposed.
One may actually solve the deflated equation (3.11) without imposing any
constraint and simply apply $P_R$ to the calculated solution
at the end of the computation.

The full quark propagator $S(x,y)$ can be similarly split into
two parts,
\equation{
  S(x,y)=P_RS(x,y)+
  \sum_{k,l=1}^N\dmode_k(x)(\Zop^{-1})_{kl}\dmode_l(y)^{\dagger},
  \enum
}
the second term being the contribution along the deflation subspace
while the first coincides with
the Green function of the deflated system (3.11). In practice
eq.~(3.12) may be a starting point for the application of 
variance reduction methods such as those described in 
refs.~[\ref{LowModeGiusti},\ref{AllToAllTrinity}].

\subsection 3.2 Deflation efficiency

Some insight into why deflation is potentially beneficial 
is obtained by noting that the deflated operator
\equation{
  \Dhat=P_LD=P_LD(1-P)
  \enum
}
acts in the orthogonal complement $\subsp^{\perp}$ of the deflation subspace.
Moreover, a little algebra shows that $\Dhat$
is the Schur complement of $D$ with respect to $\subsp$ 
and that its inverse in $\subsp^{\perp}$ is given by
\equation{
  \Dhat^{-1}=(1-P)D^{-1}(1-P).
  \enum
}
The condition number of the deflated system (3.11) is thus
expected to be significantly smaller than the condition number 
of the full system 
if the low modes of the Dirac operator are
sufficiently suppressed by the projector $1-P$. 

For any given normalized quark field $\psi(x)$, the deficit
\equation{
  \epsilon=\|(1-P)\psi\|^2
  \enum
}
provides a practical measure of how well the field 
is approximated by the
deflation subspace. Useful subspaces will have to be such that
all low quark modes (in, say, the range considered in sect.~2) have small 
deficits $\epsilon$. However, contrary to what may be presumed,
the construction of such subspaces does not require 
the low modes to be computed to any accuracy
(see sect.~5).

\section 4. Domain-decomposed subspaces

The deflation subspaces considered in the following 
are based on a division of the lattice into non-overlapping
rectangular blocks of lattice points. 
Domain decompositions of this kind were previously
introduced for the Schwarz preconditioning of 
the Dirac operator and the HMC algorithm [\ref{SchwarzII},\ref{SchwarzIII}],
but the subspaces constructed in this paper are not
linked to the Schwarz preconditioning and can be used in 
many different ways.

\subsection 4.1 Block projection method

Once the lattice is divided into blocks,
local deflation subspaces may be defined
by specifying $\Nm$ orthonormal quark fields 
$\dmode^{\Lambda}_l(x)$, $l=1,\ldots,\Nm$, 
on each block $\Lambda$. 
The full deflation subspace is then spanned by the 
set of all these local subspaces and thus has dimension $N=N_b\Nm$,
where $N_b$ denotes the number of blocks in the lattice. In particular,
at fixed block size, the total number of basis fields
scales proportionally to the lattice volume $V$.

Subspaces of this kind fit the general framework discussed in the previous
section if the basis fields are relabelled 
by an index $k$ running from $1$ to $N$. 
The little Dirac operator, the deflation projectors and the deflated 
Dirac operator are thus defined as before.
An obvious advantage of the construction is that the 
application of the projector $P$ to a given quark field $\psi(x)$,
\equation{
   P\psi(x)=\sum_{\Lambda}\sum_{l=1}^{\Nm}
   \dmode^{\Lambda}_{l}(x)\left(\dmode^{\Lambda}_{l},\psi\right),
   \enum
}
requires a number of arithmetic operations proportional to the lattice
volume times $\Nm$ (rather than $N$).
From the point of view of the operations count and the memory 
requirements,
the subspace thus behaves as if it were spanned 
by only $\Nm$ fields.
A notable exception to this rule is the little Dirac operator, which
always acts in a space of dimension $N$.

\subsection 4.2 Building domain-decomposed subspaces from global fields

In practice the block fields 
$\dmode^{\Lambda}_l(x)$, $l=1,\ldots,\Nm$,
will be obtained starting from a set of $\Nm$ globally
defined quark fields $\psi_l(x)$. The procedure is very simple
and begins by projecting the input fields to the blocks, i.e.~by
defining the fields
\equation{
  \psi^{\Lambda}_l(x)=\cases{\psi_l(x)& if $x\in\Lambda$,\cr
                             \noalign{\vskip1ex}
                                  0   & otherwise,\cr}
  \enum
}
on each block $\Lambda$. 
The Gram--Schmidt orthonormalization process
is then applied to these and the orthonormalized fields
are taken to be the basis elements $\dmode^{\Lambda}_l(x)$.

The subspace generated in this way contains 
the fields $\psi_l(x)$, but since the number of basis fields is multiplied
by the number of blocks, the subspace tends to be much larger
than the space spanned by the input fields.

\subsection 4.3 Deflation of the free-quark theory

For illustration and in order to motivate the further developments,
it is now helpful to briefly consider the case of the free-quark theory.
As will become clear below, a good choice of the basis
fields in this theory are the constant modes.
Since the quark fields carry a Dirac and a colour index,
one has $\Nm=12$ orthonormal constant modes on each block.

If periodic or anti-periodic boundary conditions are imposed,
the eigenmodes of the Dirac operator are plane waves of the form
\equation{
  \psi_p(x)=u_p\,\rme^{ipx},
  \enum
} 
where $u_p$ is a spinor that depends on the momentum $p$
but not on the position $x$.
Assuming an
$L^4$ lattice and a block division into blocks of size $b^4$
(where $L$ is an integer multiple of $b$),
a straightforward computation then shows that
\equation{
  \|(1-P)\psi_p\|^2=\epsilon_p\|\psi_p\|^2,
  \qquad
  \epsilon_p=
  \frac{1}{12}p^2\left(b^2-a^2\right)+\rmO(p^4b^4).
  \enum
}
The projection to the orthogonal complement of the specified
deflation subspace thus suppresses the low-momentum modes
by a factor proportional to $p^2$ (see fig.~2).

\topinsert
\vbox{
\vskip0.0cm
\epsfysize=5.0cm\hskip2.2cm\epsfbox{plots/sine.eps}%
\vskip0.2cm
\figurecaption{%
Approximation of a plane wave by a superposition of 
constant block modes. In the free-quark theory,
piecewise constant deflation modes achieve high
deflation efficiencies up to momenta $p$ on the order
of the inverse of the block size $b$.
}
\vskip0.0cm
}
\endinsert

A second and perhaps more important observation is that
the deflation efficiency does not depend on the
lattice size. Even on very large lattices, all low modes
with momenta $p$ of magnitude up to some fraction of $1/b$ are 
deflated with small deficits $\epsilon_p$. 
Figure~2 also illustrates the 
fact that high deflation efficiencies can be achieved
by subspaces of fields that are only piecewise smooth,
i.e.~fields that are far from being approximate
eigenmodes of the Dirac operator.

\section 5. Local coherence and subspace generation

The discussion in the previous section suggests that
the $V^2$--problem can perhaps be solved 
using domain-decomposed deflation subspaces. 
However, no general prescription was given so far of how to choose 
the fields $\psi_l(x)$, $l=1,\ldots,\Nm$, from
which these subspaces are built (cf.~subsect.~4.2).
Such a prescription will now be developed,
based on a property of the low quark modes referred to as 
local coherence.

\subsection 5.1 Smoothness \& local coherence

In the free-quark theory,
the block projection method works out because the low-momentum modes 
are smooth on the scale of the block size $b$. 
The intuitive picture that goes along with this explanation
is rather appealing but may be difficult to carry over to the full theory.
In particular, the notion of smoothness
ceases to have a well-defined meaning
in presence of a non-trivial lattice gauge field.

A related concept which is better adapted to 
the situation in the full theory is local coherence.
Loosely speaking, a set of quark fields
is referred to as locally coherent if 
the fields are locally well approximated by a relatively small
number of fields. When projected to the blocks of a 
block lattice, for example, such fields are contained in small 
subspaces of block fields,
up to small deficits that depend on the block
size and the dimension of the local subspaces. 

It is quite clear that the block projection method can only work out 
if the low quark modes are locally coherent in this sense.
Whether this is so appears to be difficult to tell 
on the basis of simple reasoning alone.
The free-quark theory certainly provides little 
guidance at this point, because the physics of the low modes is 
completely different from the one in the full theory.

\subsection 5.2 Numerical experiments

Local coherence is a property that can be investigated numerically
in a straightforward manner. One begins with an accurate
computation of the low-lying eigenvalues and associated
eigenmodes of $D^{\dagger}D$ and constructs a domain-decomposed 
subspace from an arbitrary subset of the calculated modes, 
following the lines of subsect.~4.2.
The question is then whether all other low modes 
are also contained in this subspace, up to small deficits $\epsilon$.

Several numerical experiments of this kind 
were performed in two-flavour QCD
on the lattices specified in subsect.~2.1. The results 
are quite impressive and unambiguously show
that the low modes in this theory are 
locally coherent to a high degree. Moreover, the 
property appears to hold for every individual
gauge-field configuration and not just on average.

If the $64\times32^3$ lattice is divided into blocks of size
$4^4$, for example, 
and if $12$ eigenmodes out of $48$ are selected for the 
construction of the domain-decomposed subspace,
the remaining $36$ modes turn out to lie in
the subspace up to deficits $\epsilon$ ranging from
$0.03$ to $0.06$. The deficits increase with the block size, 
but become smaller if more modes are used for the subspace construction.
On the $48\times24^3$ lattice the situation is 
practically the same, i.e.~similar deficits are obtained
for a given block size and subspace dimension.

\subsection 5.3 Subspace generation

As explained in subsect.~4.2, 
the deflation subspaces constructed in this
paper are obtained by restricting a set of quark fields
$\psi_l(x)$, $l=1,\ldots,\Nm$, to the blocks of
a block division of the lattice.
The fields could be taken to be low eigenmodes of 
the Dirac operator, but it is far more economical to 
generate them by a relaxation process, starting from a set of 
random fields.

A relaxation method that can be used in this context
is inverse iteration, where
the fields are updated a number of times according to 
\equation{
   \psi_{l}(x)\to\hbox{``}D^{-1}\kern1pt\hbox{''}\psi_{l}(x),
   \quad
   l=1,\ldots,\Nm.
   \enum
}
The inverse of the Dirac operator is put in quotes in this formula,
because an accurate solution of the Dirac equation is 
not required.
The application of a few cycles of the Schwarz alternating
procedure [\ref{SchwarzII}], for example, actually already has
the desired relaxation effect. Moreover, the procedure
can be bootstrapped by using the current set of fields
to deflate the Dirac equation and thus to accelerate the 
approximate solution of the equation in the next step
(see sect.~6).

Inverse iteration rapidly depletes the 
components of the fields parallel to
the high modes of the Dirac operator.
After a few cycles, 
the fields then satisfy the bound
\equation{
  \|D\psi_l\|\leq M\|\psi_l\|,    
   \quad l=1,\ldots,\Nm.
  \enum
} 
for some value of $M$ in the range of the low eigenvalues
of $(D^{\dagger}D)^{1/2}$. 

An important remark is now
that such fields are, to a good approximation,
linear combinations of the low quark modes. They are therefore
locally coherent with these and consequently generate domain-decomposed 
subspaces that approximate the low modes up to small deficits.
Some further experimenting actually confirms this and 
also shows that the deflation efficiencies are not very different
from those achieved by domain-decomposed subspaces 
built from exact low modes.

\subsection 5.4 Choice of parameters

The deflation efficiency of the subspaces generated in this way
depends on the block size, the dimension $\Nm$ of the local subspaces
and on the number and quality of inverse iteration steps 
that were applied.
Choosing small blocks and large numbers $\Nm$ of fields results in 
high deflation efficiencies but tends to increase the
condition number of the little Dirac operator and thus 
the computer time required for the application of the oblique
projectors $P_L$ and $P_R$. Similarly, the beneficial
effects of high numbers of fields and inverse iteration steps 
must be balanced against the effort spent for the subspace generation.

On the lattices specified in subsect.~2.1, choosing blocks of size 
$4^4$ and setting $\Nm=20$ turns out to be a good compromise.
Highly efficient deflation subspaces are obtained in this case if 
the relaxation procedure is stopped when 
the bound (5.2) is satisfied
for a value of $M$ in the $\overline{\rm MS}$ scheme
equal to $100$ MeV or so (cf.~sect.~2). This level is
reached after $11$ inverse iteration steps and requires 
a computational effort equivalent to about $190$ applications 
of the Dirac operator per field
(if slightly less effective deflation subspaces are acceptable,
one can do with $8$ steps and $130$ applications).

In general the parameters will have to be tuned empirically.
If a deflated solver program like the one described in the next section
is available, the deflation efficiency of a given subspace can be 
quickly determined by measuring the time required for the solution
of the Dirac equation to a specified accuracy. Computations of 
the low quark modes are then again not required.
The inverse iteration steps can, incidentally, be carried out
at a valence quark mass different from the sea-quark mass.
For reasons of efficiency, it is in fact recommended to set the bare mass
in this process
to a value close to (or even equal to) the critical mass.

\section 6. Deflation-accelerated solver for the Dirac equation

Low-mode deflation is expected to be useful in several areas of
lattice QCD, 
some of which [\ref{NeffEtAl}--\ref{LowModeKpipi}]
were already mentioned in sect.~1. The principal goal in this
section is to show, in
a concrete case, that the deflation subspaces constructed following 
the prescriptions given in the 
previous section are very effective and that they actually 
do provide a solution to the $V^2$--problem.

\subsection 6.1 Preconditioned Dirac equation

Once the deflation subspace is generated, the deflated Dirac equation
(3.11) can be solved straightforwardly using any of the well-known
Krylov space algorithms (see ref.~[\ref{Saad}], for example).
However, from the point
of view of the execution time, such a solver may not perform
too well, because the little system
\equation{
  \sum_{l=1}^N\Zop_{kl}v_l=w_k,
  \qquad k=1,\ldots,N, 
  \enum
}
must be solved, for one source vector $w=(w_1,\ldots,w_N)$,
each time the projector $P_L$ is applied.
As explained in appendix A, there are efficient algorithms to 
solve the little Dirac equation, but the computational effort
remains non-negligible.

A better balance of deflation and other operations
can be achieved by right-pre\-con\-ditioning the deflated equation.
The solver discussed in the following includes
the Schwarz preconditioner $M_{\rm sap}$ introduced 
in ref.~[\ref{SchwarzII}], but a
polynomial preconditioner or a fixed number
of GMRES iterations [\ref{GMRESRI},\ref{GMRESRII}]
may do just as well. In the case of the Schwarz preconditioner, the
preconditioned deflated equation reads
\equation{
  P_LDM_{\rm sap}\phi(x)=P_L\eta(x)
  \enum
}
and the solution of eq.~(3.11) is then given by 
$\chi(x)=P_RM_{\rm sap}\phi(x)$. The important point to note is
that the preconditioning reduces the iteration count of the Krylov space 
algorithm and thus the overhead generated by the deflation projector.

\subsection 6.2 Krylov space solver and the deflation-relaxation interplay

Both the Schwarz preconditioner and the deflation projector
involve approximate iterative procedures.
The GCR algorithm is a recommended Krylov space solver
in this situation, because it allows for inexact
preconditioning without compromising the correctness
of the solution (see ref.~[\ref{Saad}] for a general
discussion and ref.~[\ref{SchwarzII}] for a description of
the algorithm in the context of lattice QCD).

An interesting feature of the GCR algorithm is that the
Krylov space is extended, in each step, in a direction
$\xi(x)=M_{\rm sap}\rho(x)$ where $\rho(x)$ denotes the current residue.
The latter satisfies $P_L\rho(x)=\rho(x)$ by construction and
is therefore orthogonal to the deflation subspace (cf.~sect.~3).
When acting on such a field, the alternating Schwarz procedure
(which is basically a relaxation method) tends to be quite effective
in producing an approximate solution of the Dirac equation 
$D\xi(x)=\rho(x)$. Low-mode deflation thus has the effect of 
improving the efficiency of the preconditioner. 

Once $\xi(x)$ is calculated, the minimal residue 
in the so extended Krylov space is determined
by computing $P_LD\xi(x)$ and by applying an orthogonalization 
process. There is thus an interplay between deflation and
relaxation, where the low-mode and the high-mode 
components of the residue are reduced in alternation by
the deflation projector and the Schwarz preconditioner.

\topinsert
\newdimen\digitwidth
\setbox0=\hbox{\rm 0}
\digitwidth=\wd0
\catcode`@=\active
\def@{\kern\digitwidth}
\tablecaption{Average solver iteration numbers $N_{\rm X}$
and executing times $t\,^*$} 
\vskip-0.5ex
$$\vbox{\settabs\+&%
                  xxxxxxxxxx&x&
                  xxxxxxx&&
                  xxxxxxxxxxxxxxxxxx&x&
                  xxxxxxxxxxxx&x&
                  xxxxxxxxxxxxxxxxxx&x&\cr
\thicktablerule
\vskip1.0ex
                \+& 
                 && 
                 && \hfill EO+BiCGstab\hbox{\hskip0.1em}\hfill
                 && \hfill SAP+GCR\hfill
                 && \hfill DFL+SAP+GCR\hbox{\hskip0.1em}\hfill
                 &\cr
\vskip0.0ex
                \+& \hfill Lattice\hfill
                 && \hfill $\kval$\hfill
                 && \hfill $N_{\rm BiCG}$\hskip1.3em$t\,$[sec]\hskip1.4em
                 && \hfill $N_{\rm GCR}$\hskip1.3em$t\,$[sec]\hskip0.4em
                 && \hfill $N_{\rm GCR}$\hskip1.3em$t\,$[sec]\hskip1.5em
                 &\cr
\vskip1.0ex
\thintablerule
\vskip1.2ex
  \+& \hfill$48\times24^3$\hfill
  &&  \hfill $0.13550$\hfill
  &&  \hfill $314$ \hskip1.5em $@57$\hskip2.0em
  &&  \hfill $ 50$ \hskip1.7em $@35$\hskip0.7em
  &&  \hfill $ 17$ \hskip2.0em $15$\hskip2.2em &\cr
\vskip0.3ex
  \+& $$\hfill
  &&  \hfill $0.13590$\hfill
  &&  \hfill $492$ \hskip1.5em $@90$\hskip2.0em 
  &&  \hfill $ 78$ \hskip1.7em $@56$\hskip0.7em 
  &&  \hfill $ 19$ \hskip2.0em $18$\hskip2.2em &\cr
\vskip0.3ex
  \+& $$\hfill
  &&  \hfill $0.13610$\hfill
  &&  \hfill $684$ \hskip1.5em $125$\hskip2.0em 
  &&  \hfill $110$ \hskip1.7em $@78$\hskip0.7em 
  &&  \hfill $ 20$ \hskip2.0em $19$\hskip2.2em &\cr
\vskip0.3ex
  \+& $$\hfill
  &&  \hfill $0.13625$\hfill
  &&  \hfill $954$ \hskip1.5em $174$\hskip2.0em 
  &&  \hfill $157$ \hskip1.7em $118$\hskip0.7em 
  &&  \hfill $ 21$ \hskip2.0em $21$\hskip2.2em &\cr
\vskip0.3ex
  \+& $$\hfill
  &&  \hfill $0.13635$\hfill
  &&  \hfill $1269$ \hskip1.5em $231$\hskip2.0em 
  &&  \hfill $ 227$ \hskip1.7em $170$\hskip0.7em 
  &&  \hfill $  22$ \hskip2.0em $22$\hskip2.2em &\cr
\vskip2.0ex
\thintablerule
\vskip1.2ex
  \+& \hfill$64\times32^3$\hfill
  &&  \hfill $0.13550$\hfill
  &&  \hfill $323$ \hskip1.5em $@72$\hskip2.0em
  &&  \hfill $ 52$ \hskip1.7em $@45$\hskip0.7em
  &&  \hfill $ 17$ \hskip2.0em $20$\hskip2.2em &\cr
\vskip0.3ex
  \+& $$\hfill
  &&  \hfill $0.13590$\hfill
  &&  \hfill $520$ \hskip1.5em $115$\hskip2.0em 
  &&  \hfill $ 83$ \hskip1.7em $@71$\hskip0.7em 
  &&  \hfill $ 20$ \hskip2.0em $23$\hskip2.2em &\cr
\vskip0.3ex
  \+& $$\hfill
  &&  \hfill $0.13610$\hfill
  &&  \hfill $748$ \hskip1.5em $165$\hskip2.0em 
  &&  \hfill $120$ \hskip1.7em $103$\hskip0.7em 
  &&  \hfill $ 21$ \hskip2.0em $26$\hskip2.2em &\cr
\vskip0.3ex
  \+& $$\hfill
  &&  \hfill $0.13625$\hfill
  &&  \hfill $1125$ \hskip1.5em $248$\hskip2.0em 
  &&  \hfill $183$ \hskip1.7em $171$\hskip0.7em 
  &&  \hfill $ 23$ \hskip2.0em $29$\hskip2.2em &\cr
\vskip0.3ex
  \+& $$\hfill
  &&  \hfill $0.13635$\hfill
  &&  \hfill $1663$ \hskip1.5em $366$\hskip2.0em 
  &&  \hfill $ 294$ \hskip1.7em $267$\hskip0.7em 
  &&  \hfill $  25$ \hskip2.0em $32$\hskip2.2em &\cr
\vskip1.2ex
\thicktablerule

\vskip1.5ex
\+{\footnotefont $^*$ Using $24$ and $64$ processors, respectively, in
the case of the $48\times24^3$ and the $64\times32^3$ lattice}\cr
}
$$
\vskip-2.0ex
\endinsert

\subsection 6.3 Performance tests

The performance of the complete algorithm
(DFL+SAP+GCR for short) was determined on the lattices specified in 
subsect.~2.1, at the values of the (valence) quark mass 
that correspond to the hopping 
parameters $\kval$ listed in table~1.
In this range of masses,
the bare current-quark mass $\mval$ decreases from
the strange-quark mass $m_s$ to approximately $\frac{1}{6}m_s$
[\ref{QCDlite}], where the condition number of the Dirac operator 
reaches a value of about $1900$. 

In order for the effects of low-mode deflation to be clearly seen,
the performance measurements were extended to the 
even-odd preconditioned BiCGstab algorithm (EO+BiCGstab)
[\ref{BiCGstabI},\ref{BiCGstabII}]
and the Schwarz-preconditioned GCR algorithm without deflation
(SAP+GCR) [\ref{SchwarzII}].
In all cases, the tests consisted in measuring 
the solver iteration numbers and the computer time
required for the solution of the 
full Dirac equation (3.9) to a precision where
$\|\eta-D\psi\|\leq10^{-10}\|\eta\|$.
Timings were taken on a recent PC cluster with Infiniband network,
using $12$ and $32$ double-processor nodes for the tests on
the $48\times24^3$ and the $64\times32^3$ lattice respectively. Only
highly optimized, parallel programs were used that include machine-specific
enhancements such as those mentioned in ref.~[\ref{SchwarzII}].
Quoted solver iteration numbers and timings are averages
over $50$ gauge-field configurations.

The algorithmic parameters were set to the same values on the 
$48\times24^3$ and the $64\times32^3$ lattice. In particular,
the deflation subspaces were constructed by applying
$11$ inverse iteration steps to $\Nm=20$ random fields
and by projecting them to a division of the lattice into 
blocks of size $4^4$.
In the case of the Schwarz preconditioner,
the block size was taken to be $8\times4^3$ and all other 
parameters were set to the standard values previously used
in refs.~[\ref{SchwarzII},\ref{SchwarzIII},\ref{QCDlite}].
A fairly small value, $\Nkv=16$, turned out to be a
satisfactory choice for the maximal number $\Nkv$ of 
Krylov vectors that may be generated
before the GCR algorithm is restarted (larger values, up to $\Nkv=32$, 
had to be used in the case of the SAP+GCR solver).

\topinsert
\vbox{
\vskip0.0cm
\epsfxsize=9.0cm\hskip1.5cm\epsfbox{plots/perf32.eps}
\vskip0.3cm
\figurecaption{%
Average execution time $t$ needed for the solution of 
the lattice Dirac equation on the $64\times32^3$ lattice
as a function of the bare valence quark mass $m_{\rm val}$
given in units of the lattice spacing $a$.
The lattice, algorithm and test parameters are
as specified in subsects.~2.1 and 6.3.
Dotted lines are drawn to guide the eye.
}
\vskip0.0cm
}
\endinsert

As is evident from the test results quoted in table~1, low-mode
deflation significantly reduces both the solver iteration
numbers and 
the time needed to solve the Dirac equation to a specified precision.
Particularly impressive is the fact, illustrated in fig.~3,
that the deflated algorithm has a flat scaling behaviour with
respect to the quark mass. Moreover, 
the solver iteration numbers on the two lattices are nearly the same,
which is very much in line with the expectation that 
the efficiency of the domain-decomposed deflation subspaces 
is independent of the lattice volume and that they
provide a solution to the $V^2$--problem. 

Contrary to the solver iteration numbers, the timings quoted in 
the last column of table~1 
are sensitive to the time required
for the application of the deflation projector $P_L$ and thus
to the average time needed for the solution of the little Dirac equation
(see appendix A). The application of the projector actually 
consumed as much as
$25$\% of the total time on the small lattice and up to 
$30$\% on the big lattice.

\subsection 6.4 Miscellaneous remarks

(1) {\it Partially quenched QCD}. 
In the tests reported in the previous subsection, 
the deflation subspace was generated only once per gauge-field 
configuration, i.e.~the same subspace was used
at all values of the valence-quark mass considered.

\vskip1ex\noindent
(2) {\it Deflation overhead}. The average time spent for
the subspace generation was $150$ and $184$ seconds, respectively,
on the $48\times24^3$ and the $64\times32^3$ lattice. These figures
include the time needed for the computation of the 
little Dirac operator ($3.3$ and $3.9$ seconds).
The computational effort required for the preparatory work
thus becomes quickly negligible if several quark propagators 
are to be computed.

\vskip1ex\noindent
(3) {\it Solver stability}.
In the case of the deflated solver, the GCR iteration numbers
$N_{\rm GCR}$ tend to be very stable. 
The iteration
numbers observed in the tests actually deviated by at most
$1$ from their average values, 
except at
the smallest quark mass on the $64\times32^3$ lattice, where the maximal 
value of $N_{\rm GCR}$ ever seen was $27$. 

\vskip1ex\noindent
(4) {\it Acceleration of the HMC algorithm}.
The HMC simulation algorithm [\ref{HMC}]
requires the lattice Dirac equation
to be solved at regular intervals along the
trajectories in field space which lead from the current to the next 
configuration. Whether the use of low-mode
deflation is profitable in this case depends on the quark mass and
the precision requirements.

On the lattices specified in subsect.~2.1, for example,
an acceleration is achieved at hopping parameters 
$\kappa_{\rm sea}\geq0.13625$, if
the relative solver tolerance is set to $10^{-7}$ or less and
if, say, $8$ inverse iteration steps are used for the subspace 
generation.
At these fairly small quark masses,
the scaling behaviour of the HMC algorithm 
is then softened by nearly one power in the quark mass.

In practice much larger speed-up factors can conceivably be obtained
by updating the deflation subspace along the trajectories in field space
rather than generating the subspace from scratch each time
the Dirac equation must be solved.
Moreover,
starting from the exact factorization
\equation{
   \det D=\det\Zop\det\hat{D}
   \enum
}
of the quark determinant, the HMC algorithm itself can perhaps be deflated 
too, in which case further accelerations and an improved stability 
of the algorithm will presumably be achieved.

\section 7. Concluding remarks

An important qualitative result of this paper is 
the demonstration that the low quark modes can be 
simultaneously deflated using local subspaces of low dimension.
Some further clarification (an analytic proof of the local
coherence of the low modes, for example)
would certainly be welcome, but the numerical studies conducted so far
leave little doubt that the construction does indeed provide a solution 
to the $V^2$--problem.

Variance reduction methods, such as low-mode averaging 
[\ref{LowModeGiusti}]
and all-to-all propagator techniques [\ref{BaliEtAl},\ref{AllToAllTrinity}], 
will probably be able to profit from these developments.
The performance of the deflation-accelerated solver 
for the lattice Dirac equation discussed in the previous section
is, in any case, quite impressive, particularly so at the 
smallest quark masses considered. In many cases the computational
effort required for the calculation of hadronic correlation 
functions is thus significantly reduced.

The possible inclusion of deflation ideas in QCD simulation 
algorithms is an intriguing perspective, since this may allow simulations
close to the physical values of the light-quark masses
to be performed with an effort not very much larger 
than the one required at a sea-quark mass equal to, say,
a fourth of the physical strange-quark mass.

\vskip1.0ex
I am indebted to Leonardo Giusti and Peter Weisz for critical
comments on a first version of the paper. The gauge-field configurations
used for the numerical studies reported in this paper
were generated by the authors of ref.~[\ref{QCDlite}]. 
All computations were performed on a 
dedicated PC cluster at CERN and on a CRAY XT3 at the Swiss
National Supercomputing Centre (CSCS). I am grateful to these
institutions for providing the required computer resources.

\appendix A. Solution of the little Dirac equation

In practice the dimension $N$ of the domain-decomposed 
deflation subspaces introduced in this paper tends to be
so large that an exact solution of the little Dirac equation
(6.1) is not a viable option. The iterative solver proposed here 
is based on even-odd preconditioning,
global-mode deflation and the GCR algorithm.

\subsection A.1 Computation of the little Dirac operator

The block division of the lattice implies a decomposition of 
the little Dirac operator into $\Nm\times \Nm$ block
matrices $\ZopB_{\Lambda\Lambda'}$, whose matrix elements
are given by
\equation{
   (\ZopB_{\Lambda\Lambda'})_{kl}=
   \bigl(\dmode^{\Lambda}_k,D\dmode^{\Lambda'}_l\bigr),
   \qquad k,l=1,\ldots,\Nm.
   \enum
}
Since the Wilson--Dirac operator has only nearest-neighbour
hopping terms, most of these matrices vanish and a moment
of thought reveals that
the little Dirac operator actually couples nearest-neighbour 
blocks only (see fig.~4).

\topinsert
\vbox{
\vskip0.0cm
\epsfxsize=4.0cm\hskip4.0cm\epsfbox{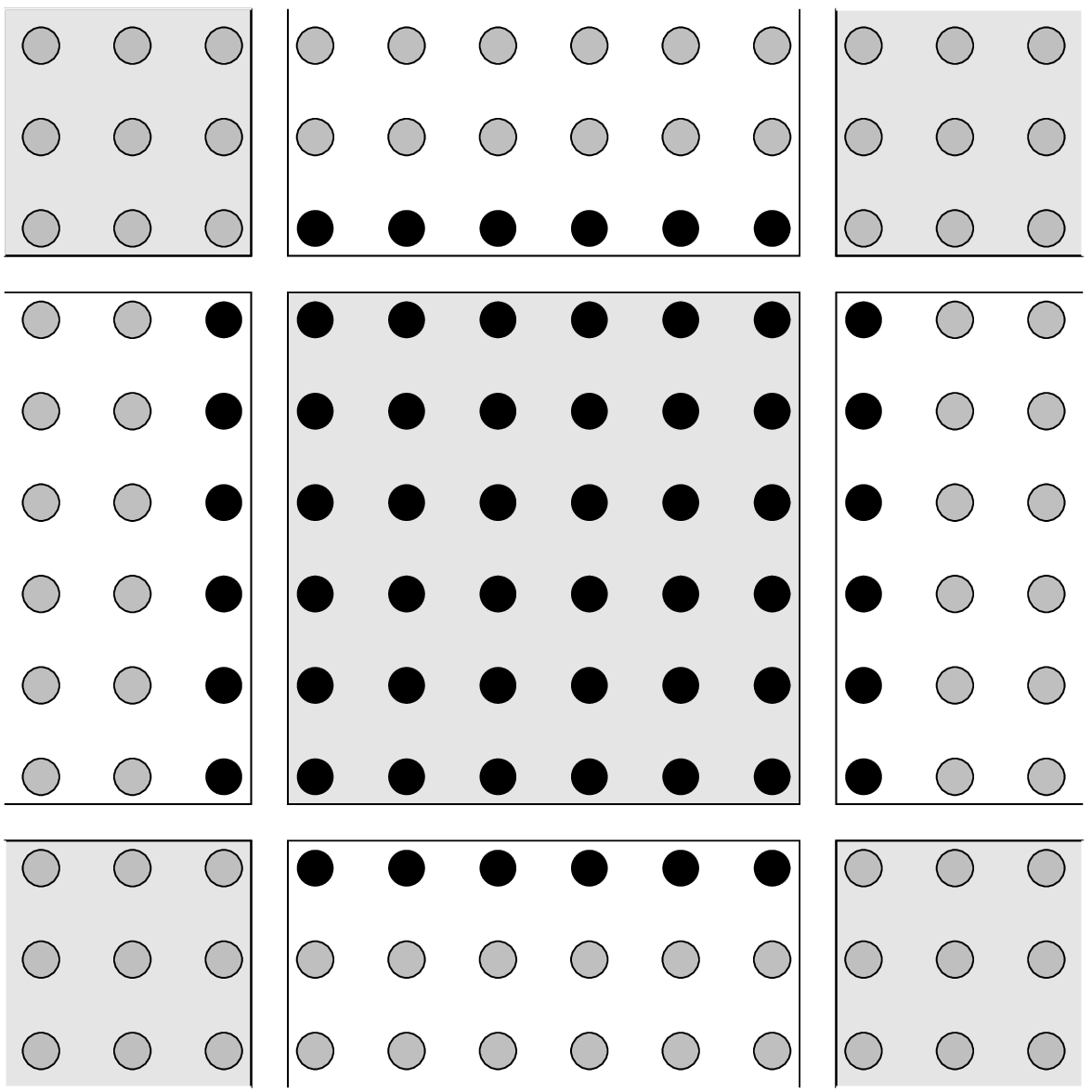}
\vskip0.3cm
\figurecaption{%
Support of the function $D\phi(x)$ (black points) if $\phi(x)$ 
is supported on the grey block in the centre of the figure. 
In particular, the matrix elements (A.1) vanish unless the blocks
$\Lambda$ and $\Lambda'$ coincide or are nearest neighbours.
}
\vskip0.0cm
}
\endinsert

The computation of the scalar products (A.1) is straightforward
and requires a total effort proportional to the lattice volume times
$\Nm^2$. Note, however, that the operations count tends to increase 
rapidly if lattice Dirac operators
with hopping terms extending over two or more links are considered.

\subsection A.2 Even-odd preconditioning

The even-odd preconditioning
familiar from the full lattice Dirac operator can also be applied
to the little Dirac operator in its block form if 
there is an even number of blocks in each direction (which 
is here assumed to be the case).
If the so-called symmetric preconditioning is chosen [\ref{MsJLQCD}],
the block matrices representing the 
preconditioned operator on even blocks $\Lambda,\Lambda'$ are
given by
\equation{
  \hat{\ZopB}_{\Lambda\Lambda'}=
  \delta_{\Lambda\Lambda'}-\sum_{\Omega}
  (\ZopB_{\Lambda\Lambda})^{-1}\ZopB_{\Lambda\Omega}(\ZopB_{\Omega\Omega})^{-1}
  \ZopB_{\Omega\Lambda'},
  \enum
}
where the sum extends over the common neighbours $\Omega$ 
of the blocks $\Lambda$ and $\Lambda'$. 

The matrices (A.2)
do not need to be stored in the memory of the computer,
because the action of the preconditioned operator on
a complex field can be computed in two steps, first hopping
from the even to the odd blocks and then back to the even blocks.
Some work can however be saved by storing the 
matrices $(\ZopB_{\Lambda\Lambda})^{-1}\ZopB_{\Lambda\Lambda'}$
for all nearest-neighbour pairs $\Lambda,\Lambda'$ of blocks.

On physically small blocks $\Lambda$, the diagonal 
block matrices $\ZopB_{\Lambda\Lambda}$ tend to be safely invertible,
but the program should check this and return
to the original system if an ill-conditioned matrix 
is encountered (this never happened in the tests
reported in this paper).

\subsection A.3 Global-mode deflation

As explained in subsect.~4.2, the basis fields $\dmode^{\Lambda}_l(x)$
on the blocks $\Lambda$ are obtained starting 
from a set of global fields $\psi_l(x)$, $l=1,\ldots,\Nm$. 
The latter span a subspace in the generated
deflation subspace which may be used to deflate
the little Dirac operator. Actually only the components of the 
global fields on the even blocks are used to build this ``little
deflation subspace'', because the little Dirac equation is
to be deflated in its even-odd preconditioned form.

The equation is deflated 
following the general procedures described in sect.~3.
One simply has to replace the full Dirac operator 
by the even-odd preconditioned little Dirac operator
and the quark fields by complex fields with $N/2$ components.
Note that the 
``little little Dirac operator'' is an $\Nm\times \Nm$ matrix
that can be inverted to machine precision with a negligible effort.

Global-mode deflation is straightforward to 
implement and tends to reduce the condition number of the little
system quite significantly (by about a factor $3$ in the cases studied
so far).

\subsection A.4 Solver performance

Similarly to the full system, the deflated preconditioned little equation
can be solved using the GCR algorithm. 
Tests of the complete solver were then performed using the same
subspaces as in subsect.~6.3.
In particular, the number of GCR iterations
$N_{\rm GCR}$ and the time $t$ needed to solve the little
equation to a relative precision of $10^{-12}$ were determined
and are quoted in columns $3$ and $4$ of table~2.

\topinsert
\newdimen\digitwidth
\setbox0=\hbox{\rm 0}
\digitwidth=\wd0
\catcode`@=\active
\def@{\kern\digitwidth}
\tablecaption{GCR iteration numbers and time$\,^*$ needed for the solution
of the little system} 
\vskip-0.5ex
$$\vbox{\settabs\+&%
                  xxxxxxxxxxxx&x&
                  xxxxxxxxxxxx&xx&
                  xxxxxxxxxxxx&xx&
                  xxxxxxxxx&xx&
                  xxxxxxxxxxxx&xx&\cr
\thicktablerule
\vskip0.5ex
                \+& \hfill Lattice\hfill
                 && \hfill $\kval$\hfill
                 && \hfill $N_{\rm GCR}$\hfill
                 && \hfill $t\,$[sec]\hfill
                 && \hfill $\overline{N}_{\rm GCR}$\hfill
                 &\cr
\vskip1.0ex
\thintablerule
\vskip1.2ex
  \+& \hfill$48\times24^3$\hfill
  &&  \hfill $0.13550$\hfill
  &&  \hfill $@84$\hfill
  &&  \hfill $0.26$\hfill
  &&  \hfill $24$\hfill &\cr
\vskip0.3ex
  \+& $$\hfill
  &&  \hfill $0.13590$\hfill
  &&  \hfill $105$\hfill 
  &&  \hfill $0.32$\hfill 
  &&  \hfill $30$\hfill &\cr
\vskip0.3ex
  \+& $$\hfill
  &&  \hfill $0.13610$\hfill
  &&  \hfill $120$\hfill 
  &&  \hfill $0.37$\hfill 
  &&  \hfill $34$\hfill &\cr
\vskip0.3ex
  \+& $$\hfill
  &&  \hfill $0.13625$\hfill
  &&  \hfill $136$\hfill 
  &&  \hfill $0.42$\hfill
  &&  \hfill $38$\hfill &\cr
\vskip0.3ex
  \+& $$\hfill
  &&  \hfill $0.13635$\hfill
  &&  \hfill $150$\hfill
  &&  \hfill $0.46$\hfill
  &&  \hfill $42$\hfill &\cr
\vskip2.0ex
\thintablerule
\vskip1.2ex
  \+& \hfill$64\times32^3$\hfill
  &&  \hfill $0.13550$\hfill
  &&  \hfill $@94$\hfill
  &&  \hfill $0.37$\hfill
  &&  \hfill $27$\hfill &\cr
\vskip0.3ex
  \+& $$\hfill
  &&  \hfill $0.13590$\hfill
  &&  \hfill $126$\hfill
  &&  \hfill $0.49$\hfill
  &&  \hfill $36$\hfill &\cr
\vskip0.3ex
  \+& $$\hfill
  &&  \hfill $0.13610$\hfill
  &&  \hfill $154$\hfill
  &&  \hfill $0.60$\hfill
  &&  \hfill $44$\hfill &\cr
\vskip0.3ex
  \+& $$\hfill
  &&  \hfill $0.13625$\hfill
  &&  \hfill $188$\hfill
  &&  \hfill $0.73$\hfill 
  &&  \hfill $54$\hfill &\cr
\vskip0.3ex
  \+& $$\hfill
  &&  \hfill $0.13635$\hfill
  &&  \hfill $220$\hfill
  &&  \hfill $0.85$\hfill
  &&  \hfill $62$\hfill &\cr
\vskip1.2ex
\thicktablerule

\vskip1.5ex
\+{\footnotefont $^*$ Using $24$ and $64$ processors, respectively, in
the case of the $48\times24^3$ and the $64\times32^3$ lattice}\cr
}
$$
\vskip-2.0ex
\endinsert

The dependence of these figures on the valence quark mass
and the lattice volume is noticeable, but one can also
see that the solver iteration numbers increase
only relatively slowly towards the smaller quark masses.
In practice all these variations are not too important,
because the solution of the little system eventually 
consumes only a fraction of the time spent for the solution 
of the full system.

\subsection A.5 Using adapted precision

It is still worth including another improvement, however, which
exploits the fact that the outer GCR algorithm 
(the one that solves the full system) is restarted from time 
to time, usually when the dimension of the generated Krylov
space reaches the specified maximal value.
Before each
restart, the current residue is recomputed with high precision so
that any inaccuracies which may have accumulated during the 
last cycle do not propagate to the next cycle.

For this reason it is permissible to solve the little Dirac equation
to low precision inside the cycles of the outer algorithm. In the 
tests reported in subsect.~6.3, for example, the required 
relative tolerances were set to $10^{-6}$ and $10^{-12}$, respectively,
inside and outside the cycles of the algorithm. The average solver
iteration numbers are then practically reduced by a factor $2$.

They can actually be reduced even further by adapting the precision
as one proceeds from one Krylov vector to the next within a cycle. 
This is possible
because the GCR algorithm operates directly on the minimal residuals
in the generated Krylov spaces. Their magnitude decreases monotonically
and need to be computed essentially only to a fixed decimal precision.
The required precision for the solution of the little system can therefore be 
reduced in proportion to the norm of the quark fields on which 
the deflation projector $P_L$ acts.

Once all these improvements are installed, the average
iteration numbers $\overline{N}_{\rm GCR}$ 
required for the solution of the little system
in the course of the cycles of the outer algorithm are reduced to the 
figures quoted in the last column of table~2.
At the smallest quark mass on the $64\times32^3$ lattice, for example,
the time spent for the solution of the little system 
sums up to about $6$ seconds, i.e.~about $19$\% of the 
total time needed for the solution of the full system.

\beginbibliography


\bibitem{FrankVuik}
J. Frank, C. Vuik,
SIAM J. Sci. Comput. 23 (2001) 442

\bibitem{NabbenVuik}
R. Nabben, C. Vuik,
SIAM J. Sci. Comput. 27 (2006) 1742


\bibitem{NeffEtAl}
H. Neff, N. Eicker, T. Lippert, J. W. Negele, K. Schilling,
Phys. Rev. D64 (2001) 114509

\bibitem{LowModeGiusti}
L. Giusti, P. Hern\'andez, M. Laine, P. Weisz, H. Wittig,
JHEP 0404 (2004) 013

\bibitem{LowModeDeGrand}
T. A. DeGrand, S. Schaefer,
Comput. Phys. Commun. 159 (2004) 185

\bibitem{BaliEtAl}
G. S. Bali, H. Neff, T. D\"ussel, T. Lippert, K. Schilling (SESAM collab.),
Phys. Rev. D71 (2005) 114513

\bibitem{AllToAllTrinity}
J. Foley, K. J. Juge, A. O'Cais, M. Peardon, S. M. Ryan, J.-I. Skullerud,
Comput. Phys. Commun. 172 (2005) 145


\bibitem{deForcrand}
Ph. de Forcrand,
Nucl. Phys. B (Proc. Suppl.) 47 (1996) 228

\bibitem{MorganWilcox}
R. B. Morgan, W. Wilcox,
Nucl. Phys. (Proc. Suppl.) 106 (2002) 1067

\bibitem{LowModeKpipi}
L. Giusti, C. Hoelbling, M. L\"uscher, H. Wittig,
Comput. Phys. Commun. 153 (2003) 31


\bibitem{BanksCasher}
T. Banks, A. Casher,
Nucl. Phys. B169 (1980) 103


\bibitem{SW}
B. Sheikholeslami, R. Wohlert,
Nucl. Phys. B259 (1985) 572

\bibitem{OaImp}
M. L\"uscher, S. Sint, R. Sommer, P. Weisz,
Nucl. Phys. B478 (1996) 365


\bibitem{NPimp}
K. Jansen, R. Sommer (ALPHA collab.),
Nucl. Phys. B530 (1998) 185
[E: {\it ibid.} B643 (2002) 517]


\bibitem{QCDlite}
L. Del Debbio, L. Giusti, M. L\"uscher, R. Petronzio, N. Tantalo,
JHEP 0702 (2007) 056; {\it ibid.} 0702 (2007) 082


\bibitem{Stability}
L. Del Debbio, L. Giusti, M. L\"uscher, R. Petronzio, N. Tantalo,
JHEP 0602 (2006) 011


\bibitem{ZAalpha}
M. Della Morte et al. (ALPHA collab.),
JHEP 0507 (2005) 007


\bibitem{MsJLQCD}
S. Aoki et al. (JLQCD collab.),
Phys. Rev. D68 (2003) 054502

\bibitem{MsCPPACS}
A. Ali Khan et al. (CP-PACS collab.),
Phys. Rev. D65 (2002) 054505 [E: {\it ibid.} D67 (2003) 059901]

\bibitem{MsALPHA}
M. Della Morte et al. (ALPHA collab.),
Nucl. Phys. B729 (2005) 117


\bibitem{FukayaEtAl}
H. Fukaya et al. (JLQCD and TWQCD collab.),
Phys. Rev. Lett. 98 (2007) 172001;
{\it Two-flavor lattice QCD in the $\epsilon$-regime and
chiral random matrix theory}, arXiv:0705.3322v1 [hep-lat]


\bibitem{SchwarzII}
M. L\"uscher,
Comput. Phys. Commun. 156 (2004) 209

\bibitem{SchwarzIII}
M. L\"uscher,
Comput. Phys. Commun. 165 (2005) 199


\bibitem{Saad}
Y. Saad, Iterative methods for sparse linear systems,
2nd ed. (SIAM, Philadelphia, 2003); see also
{\tt http://www-users.cs.umn.edu/\~{}saad/}


\bibitem{GMRESRI}
H. A. van der Vorst, C. Vuik,
Num. Lin. Alg. Appl. 1 (1994) 369

\bibitem{GMRESRII}
C. Vuik,
J. Comput. Appl. Math. 61 (1995) 189


\bibitem{BiCGstabI}
H. A. van der Vorst,
SIAM J. Sci. Stat. Comput. 13 (1992) 631

\bibitem{BiCGstabII}
A. Frommer, V. Hannemann, B. N\"ockel, T. Lippert, K. Schilling,
Int. J. Mod. Phys. C5 (1994) 1073


\bibitem{HMC}
S. Duane, A. D. Kennedy, B. J. Pendleton, D. Roweth,
Phys. Lett. B195 (1987) 216

\endbibliography

\bye